\documentclass[aps,prd,13,onecolumn,preprintnumbers,floatfix,nofootinbib]{revtex4-1}
\pdfoutput=1
\usepackage{graphicx}
\usepackage{bm}
\usepackage{times}
\usepackage{slashed}
\usepackage{color}
\usepackage{slashed}
\usepackage{amsmath}
\usepackage{amsthm}
\usepackage{subfigure}

\usepackage{feynmp} 
 \usepackage{textcomp}

\newcommand{\be}{\begin{equation}}
\newcommand{\ee}{\end{equation}}
\newcommand{\bea}{\begin{eqnarray}}
\newcommand{\eea}{\end{eqnarray}}

\begin{document}

\title{Muon Anomalous Magnetic Moment in a $SU(4) \otimes U(1)_N$ Model}

\author{D. Cogollo$^{a}$}
\email{diegocogollo@df.ufcg.edu.br}

\affiliation{$^a$Departamento de F\'isica, Universidade Federal de Campina Grande,
Caixa Postal 10071, 58109-970, Campina Grande, PB, Brasil\\}

\begin{abstract}
We study the muon anomalous magnetic moment in an electroweak model based on the gauge symmetry  $SU(4)_L\otimes U(1)_N$, which has right-handed neutrinos in its spectrum, and no flavor changing neutral currents at tree level. We discuss relevant collider and electroweak constraints on the model, and derive the most stringent upper bounds on the scale of symmetry breaking based on the corrections to the muon magnetic moment. We conclude that a scale of symmetry breaking of around 2TeV might explain the muon magnetic moment anomaly. In case the anomaly is otherwise resolved, using the current and projected sensitive of g-2 experiments, we rule out scales of symmetry breaking smaller than $3.5$~TeV ($5.5$~TeV) at $1\sigma$ level.
\end{abstract}


\maketitle

\section{Introduction}
\label{intro}

Considering the astonishing agreement between its predictions and the experimental data, the Standard Model (SM) is up to now the best theory we have to understand the properties and interactions of the elementary particles. However very high precision measurements establish a strong test of the SM predictions, leading us to consider quantum corrections that cannot be disregarded more. Measuring the magnetic moments of elementary particles, especially of the muon, is one of the best examples of how to investigate new physics that can account for the experimental data.     

The Dirac equation predicts that the magnetic dipole moment $\vec{\mu}$ of a spin $\vec{S}=\dfrac{1}{2}$ particle such as the muon is given by:

\begin{equation}
\vec{\mu}=g_{\mu}\dfrac{e}{2m_{\mu}}\vec{S},
\end{equation}

with gyromagnetic ratio $g_{\mu}=2$. Loop-level corrections generate little deviations from 2 -- the anomalous magnetic moment, parametrized by $a_{\mu} = (g_{\mu}-2)/2$. Within the Standard Model framework this deviation can be precisely predicted and it is generally divided into three parts, $a_{\mu}^{SM}=a_{\mu}^{QED}+a_{\mu}^{EW}+a_{\mu}^{HAD}$ \cite{PDG}. The QED correction, which is by far the dominant contribution in the SM, includes all photonics and leptonics loops. Starting with the $\frac{\alpha}{2\pi}$ Schwinger contribution, it has been computed through 4 loops and estimated at the 5-loops\cite{5loops1,5loops2,5loops3,5loops4,5loops5,5loops6,5loops7,Hanneke:2008tm1,
Hanneke:2008tm2}. Loop contributions involving heavy $W^{\pm}$, $Z$ or the Higgs particle are labeled as $a_{\mu}^{EW}$ and are suppressed by at least a factor $\sim \frac{\alpha}{\pi}\frac{m_{\mu}^{2}}{M_{W}^{2}}$. The hadronic contributions are associated with quarks and gluons loops and can not be calculated by first principles.\\ 

The $a_{\mu}$ has been measured reaching the level of 0.54 ppm. The present difference $\Delta a_{\mu}=a_{\mu}^{exp} - a_{\mu}^{SM}=295 \pm 81 \times 10^{-11}$ yields a $3.6\sigma$ discrepancy\cite{bennet,bennet1,bennet2}, suggesting the existence of new physics that accounts for it. However the large theoretical uncertainties can overshadow the significance of this discrepancy \cite{carey}. Theoretical enhancements along with the projected experimental sensitivity for the $g-2$ experiment at Fermilab possibly reach $\Delta a_{\mu}=295 \pm 34 \times 10^{-11}$, enhancing the signal up to $5\sigma$ \cite{carey, carey1}.\\

As known, there are crucial non-addressed issues such as the neutrino masses and dark matter that motivate us to extend the SM. The $3.6\sigma$ discrepancy provides a hint of new physics too, and might be accomplished in extensions of the SM. The gauge structure of a model determines how the fermions can interact. The SM electroweak sector has the group structure $SU(2)_L \otimes U(1)_Y$, where left-handed fermionic fields are organized in doublets and the right-handed ones, except for neutrinos, are singlets by the $SU(2)_L$ symmetry. The scalar sector consists in a Higgs $SU(2)_L$ doublet, that generates masses for gauge bosons and fermions except for neutrinos. In order to have massive neutrinos and address another open questions, we must extend the SM. Compelling extensions concerning the electroweak gauge sector place the right-handed fields in the same multiplet than the left-handed ones, by means of the charge conjugation. These models have a $SU(3)_L\otimes U(1)_N$ symmetry \cite{Pisano:1991ee,Pisano:1991eee}. As this extension does not have right-handed neutrinos, there is not a ``build-in'' mechanism for generation of neutrino masses. In order to generate neutrino masses, deviations from the former proposal, where $\nu^c,\nu$ and $e$ are in the same multiplet of $SU(3)$, are required~\cite{Foot:1994ym,Foot:1994ymm}. Frameworks based on the $SU(3)_c \otimes SU(3)_L \otimes U(1)_N$ gauge symmetry, called 3-3-1 for short, can provide plausible dark matter candidates in the context of Higgs Portal \cite{331Higgsportal,331Higgsportal1,331Higgsportal2,331Higgsportal3,331Higgsportal4,
331Higgsportal5} and $Z^{\prime}$ portal \cite{Mizukoshi:2010ky,Mizukoshi:2010ky1,Mizukoshi:2010ky2,Mizukoshi:2010ky3,
Mizukoshi:2010ky4}, setting aside the direct detection controversy that has been happening regarding the low energy events \cite{Profumo:2014mpa}, possibly explaining the Galactic Center excess observed in the Fermi-LAT data \cite{Alves:2014yha,Alves:2014yha1,Alves:2014yha2}, address the dark radiation non-thermal dark matter production \cite{Kelso:2013nwa,Kelso:2013nwa1,Kelso:2013nwa2,Kelso:2013nwa3,Kelso:2013nwa4}, and even reproducing the mild $H\gamma\gamma$  excess \cite{Alves:2011kc,Alves:2011kc1,Alves:2011kc2}, among others \cite{othermotiv1,othermotiv11,othermotiv12,othermotiv13,othermotiv14,othermotiv15,
othermotiv16,othermotiv17,othermotiv18,othermotiv2,othermotiv21,othermotiv22,othermotiv23,
othermotiv24,othermotiv25,othermotiv26,othermotiv27,othermotiv28,othermotiv29,
othermotiv210,othermotiv211,othermotiv212}.

An attempt to introduce the right-handed neutrinos in a more elegant way has lead to an extension of 3-3-1 models, namely 3-4-1, where $\nu,e,\nu^c$ and $e^c$ are in the same multiplet of a $SU(4)_L\otimes U(1)_N$ electroweak theory \cite{Pisano:1994tf}. This is the kind of models we will treat here. Several works have considered the muon anomalous magnetic moment in 3-3-1 models \cite{Kelso:2013zfa,Kelso:2013zfa1,Kelso:2013zfa2}, but there is lack of results in the context of 3-4-1 frameworks. Our goal here is to assess whether this model is capable of addressing the excess reported in the muon magnetic moment with respect to the SM prediction, and derive robust bounds on the model in light of the upcoming g-2 experiment at Fermilab. We start by briefly discussing the key aspects of this model, which are relevant for the muon magnetic moment. For extended references on this model we recommend \cite{Palcu:2009ks,Palcu:2009ks1,Palcu:2009ks2,Liu:1994rx}

\section{Model $SU(4)_L  \otimes U(1)_N$}

\subsection{Electric Charge Operator}
The gauge symmetry in question is anomaly free and it yields the electric charge operator defined as \cite{Pisano:1994tf},
\begin{equation}
Q=\frac{1}{2}(\lambda_3-\frac{1}{\sqrt3}\lambda_8-\frac{2}{3}{\sqrt6}
\lambda_{15})+N,
\label{q}
\end{equation} where,

\begin{equation}
\lambda_3=diag(1,-1,0,0),\;\lambda_8=(\frac{1}{\sqrt{3}})diag(1,1,-2,0),
\;\lambda_{15}=(\frac{1}{\sqrt{6}})diag(1,1,1,-3).
\end{equation}

\subsection{Fermion Content}

In this model, we have left and right-handed charged leptons and neutrinos in the same $SU(4)_L$ multiplet, that transform as $(1,4,0)$. In order to avoid anomalies, we must have the same number of $4$ and $4^*$ multiplets, and the sum of the fermions charges must be zero. So, the quark sector consists of one generation transforming as $(3,4,+2/3)$, and the two others as $(3,4^*,-1/3)$ \cite{Pisano:1994tf}. To accomplish this, we have two new quarks $u^{\prime}$ and $J$ with charges $+2/3$ and $+5/3$ respectively, and another four $j_{2,3}$ and $d_{2,3}^{\prime}$ with charges $-4/3$ and $-1/3$,  respectively. Concerning the right-handed quarks, they are all singlets under $SU(4)_L\otimes U(1)_N$. Then, the fermion content is

\begin{equation}
f_{aL}=\left(\begin{array}{c}
\nu_a\\
l_a\\
\nu^c_a\\
l^c_a
\end{array}\right)_L \sim (1,4,0), \qquad Q_{1L} = \left(
\begin{array}{c}
u_1\\
d_1\\
u^{\prime}\\
J
\end{array}\right)_L \sim (3,4,+2/3), \qquad Q_{\alpha L} = \left(
\begin{array}{c}
j_{\alpha}\\
d_{\alpha}^{\prime}\\
u_{\alpha}\\
d_{\alpha}
\end{array}\right)_L \sim (3,4^*,-1/3),
\label{lq}
\end{equation}
where $a$ is the flavor index. As in the $SU(3)_c\otimes SU(3)_L\otimes U(1)_N$ case, the number of families $(N_f)$ must be divisible by the number of color degrees of freedom ($n$), the simplest alternative being $n=N_f=3$. Hence, $a=1,2,3$ and $\alpha=2,3$

\subsection{Scalar Sector}
In order to generate masses for the quarks the following scalar multiplets are introduced:
\begin{eqnarray}
\eta=\left(
\begin{array}{c}
\eta_1^0 \\ \eta^-_1 \\ \eta^0_2 \\ \eta^+_2
\end{array}
\right) \sim(1,4,0),
\hspace{0.5cm}
\rho=\left(
\begin{array}{c}
\rho_1^+ \\ \rho^0 \\ \rho^+_2 \\ \rho^{++}
\end{array}
\right) \sim(1,4,+1),
\hspace{0.5cm}
\chi=\left(
\begin{array}{c}
\chi_1^- \\ \chi^{--} \\ \chi^-_2 \\ \chi^0
\end{array}
\right) \sim(1,4,-1).
\label{higgs4}
\end{eqnarray}

The spontaneous symmetry breaking process occurs with $\langle\eta\rangle=(v/\sqrt{2},0,0,0)$, $\langle\rho\rangle=(0,u/\sqrt{2},0,0)$, and $\langle\chi\rangle=(0,0,0,w/\sqrt{2})$.

As for the charged leptons masses, they arise through terms as $\bar f^c_Lf_L\sim (6_A\oplus 10_S)$. A Higgs multiplet transforming as $(1,10^{\ast},0)$ has been chosen because the sextet leaves one lepton massless and the others degenerates \cite{Pisano:1994tf}:
\begin{equation}
H=\left(
\begin{array}{cccc}
H^0_1\, &\, H^+_1\, &\, H_2^0\, &\, H_2^- \\
H^+_1\, &\, H_1^{++}\, &\, H_3^+\, &\, H_3^0 \\
H^0_2\, &\, H^+_3\, &\, H^0_4\, &\, H^-_4 \\
H^-_2\, &\, H^0_3\, &\, H^-_4\, &\, H^{--}_2
\end{array}
\right) \sim (1,10^*,0).
\label{10}
\end{equation}

To preclude mixing among SM and the exotic quarks an extra multiplet $\eta'$ has been used, transforming as $\eta$, but with different vacuum expectation value (VEV), $\langle\eta'\rangle=(0,0,v'/\sqrt{2},0)$. Concerning the neutrinos masses they are generated with$\langle H_{2,3,4}^0\rangle=v''$. With those assignments $SU(4)_L\otimes U(1)_N$ breaks down to $SU(3)_L\otimes
U(1)_{N'}$ through the $\chi$ multiplet. The $SU(3)_L\otimes U(1)_{N'}$
symmetry is broken down into $U(1)_{em}$ via $\rho,\eta$, $\eta'$
and $H$ Higgs. 

The Yukawa Lagrangian is,
\begin{eqnarray}
-{\cal L}_Y&=&\frac{1}{2}G_{ab}\overline{f_{aL}^c}f_{bL}H+
F_{1k}\bar Q_{1L}u_{kR}\eta+
F_{\alpha k}\bar Q_{\alpha L}u_{kR}\rho^*\nonumber \\ & &\mbox{}
+F'_{1k}\bar Q_{1L}d_{kR}\rho+
F'_{\alpha k}\bar Q_{\alpha L}d_{kR}\eta^*+
h_1\bar Q_{1L}u'_R\eta'+h_{\alpha \beta}\bar Q_{\alpha L}d'_{\beta
R}\eta'^* \nonumber \\
& &\mbox{}
+\Gamma_1\bar Q_{1L}J_R\chi+\Gamma_{\alpha \beta}\bar Q_{\alpha
L}j_{\beta L}\chi^*+H.c.,
\label{yukawa}
\end{eqnarray}
where $a,b=e,\mu,\tau$; $k=1,2,3$; and $\alpha,\beta=2,3$, give us the following muon interactions:

\begin{eqnarray}
 \mathcal{L}_Y &\supset& -\frac{1}{2}G_{ab} \overline{f^c_{a_L}} Hf_{b_L} -\frac{1}{2}G^*_{ba} \overline{f_{b_L}} H^\dagger f^c_{a_L} \nonumber \\
                      &\supset& -\frac{G_{ab}}{2} [ \overline{\nu^c_{a_L}} (H^+_1 l_{b_L} + H^-_2 l^c_{b_L})+\overline{l^c_{a_L}} (H^+_1 \nu_{b_L} + H^{++}_1 l_{b_L} + H^+_3 \nu^c_{b_L} + H^0_3 l^c_{b_L} )  \nonumber \\
                      && + \overline{\nu_{a_L}} (H^+_3 l_{b_L} + H^-_4 l^c_{b_L}) +\overline{l_{a_L}} (H^-_2 \nu_{b_L} + H^{0}_3 l_{b_L} + H^-_4 \nu^c_{b_L} + H^{--}_2 l^c_{b_L} ) \nonumber \\
                      && -\frac{1}{2}G^*_{ba} [ \overline{\nu_{b_L}} (H^-_1 l^c_{a_L} + H^+_2 l_{a_L})+\overline{l_{b_L}} (H^-_1 \nu^c_{a_L} + H^{--}_1 l^c_{a_L} + H^-_3 \nu_{a_L} + H^0_3 l_{a_L} )\nonumber \\
                      &&
                      + \overline{\nu^c_{b_L}} (H^-_3 l^c_{a_L} + H^+_4 l_{a_L}) +\overline{l^c_{b_L}} (H^+_2 \nu^c_{a_L} + H^{0}_3 l^c_{a_L} + H^-_4 \nu_{a_L} + H^{++}_2 l_{a_L} ) ]
\end{eqnarray}

\subsection{Gauge Sector}

\subsubsection{Charged Current}

Because of the gauge group is a $SU(4)_L \otimes U(1)_N$ there are $15$ $W^i_\mu$, $i=1,...,15$ gauge bosons associated with $SU(4)_L$ and a singlet $B_\mu$ associated with $U(1)_N$. The physical charged gauge bosons $-\sqrt{2}W^+=W^1-iW^2$, $-\sqrt{2}V^-_1=W^6-iW^7$, $-\sqrt{2}V_2^-=W^9-iW^{10}$, $-\sqrt{2}V_3^-=W^{13}-iW^{14}$, $-\sqrt{2}U^{--}=W^{11}-iW^{12}$ and $\sqrt{2}X^0=W^4+iW^5$ induce the following charged current muon interactions:

\begin{eqnarray}
 \mathcal{L}_l^{CC} \supset -\frac{g}{\sqrt{2}} [ \overline{\nu} \gamma^\mu (1-\gamma_5) \mu W^+_\mu + \overline{\nu^c} \gamma^\mu (1-\gamma_5) \mu V^+_{1_\mu}
 + \overline{\mu^c} \gamma^\mu (1-\gamma_5) \nu^c V^+_{2_\mu} + \overline{\mu^c} \gamma^\mu (1-\gamma_5) \mu U^{++}_\mu]
                                 +  H.C.
\end{eqnarray}

\subsubsection{Neutral Current}
After the spontaneous symmetry breaking the neutral gauge bosons mix and have a mass matrix in the basis $W^3,W^8,W^{15},B$  given by,
\begin{equation}
\frac{g^2}{4}\left(
\begin{array}{llll}
v^2\!+\!u^2\!+\!v''^2\, &\, \frac{1}{\sqrt3}(v^2\!-\!u^2\!-\!v''^2)\, &\,
\frac{1}{\sqrt6}(v^2\!-\!u^2\!+\!2v''^2)\, &\, -2tu^2 \\
 \frac{1}{\sqrt3}(v^2\!-\!u^2\!-\!v''^2)\, &\,
\frac{1}{3}(v^2\!+\!4v'^2\!+\!u^2\!+\!v''^2)\, &
\frac{1}{3\sqrt2}(v^2\!-\!2v'^2\!+\!u^2\!-\!2v''^2)\, &\,
\frac{2}{\sqrt3}tu^2 \\
\frac{1}{\sqrt6}(v^2\!-\!u^2\!+\!2v''^2)\, &\,
\frac{1}{3\sqrt2}(v^2\!-\!2v'^2\!+\!u^2\!-\!2v''^2)
&\, \frac{1}{6}(v^2+v'^2+u^2+9w^2+4v''^2)\, &\,
\frac{2}{\sqrt6}t(u^2+3w^2) \\
-2tu^2\, &\, \frac{2}{\sqrt3}tu^2\, &\, \frac{2}{\sqrt6}t(u^2\!+\!3w^2)\,
&4t^2(u^2\!+\!w^2)
\end{array}
\right)
\label{zmass}
\end{equation} where $t\equiv g'/g$. There are four neutral gauge bosons: the massless photon, and three massive ones: $Z,Z',Z_{N}$ such that $M_Z<M_{Z'}<M_{Z_{N}}$. The lightest one is identified as the SM Z boson. In principle the diagonalization procedure has to be done numerically. However, an analytic solution can be found by setting $v=u=v''\equiv v_1$ and $v'=w\equiv v_2$, with $v_2 \gg v_1$, yielding \cite{Pisano:1994tf}
\begin{equation}
M^2_n\approx g^2 \lambda_{n} v_2^2,\quad n=0,1,2;
\label{masszs}
\end{equation}where $\lambda_{n}$ are constants given in the Appendix. The important fact is that $n=0,1,2$ refers to $Z_{N},Z$ and $Z^{\prime}$ respectively. Interestingly we find that both $Z_{N}$ and $Z^{\prime}$ give rise to sizeable contributions to the muon magnetic moment as we shall see further. For now we present the muon neutral currents,

\begin{equation}
{\cal L}_n^{NC} =-\frac{g}{2c_W}\left( \bar{l}_{L} \gamma^{\mu} l_L \alpha  + \bar{l}_{R} \gamma^{\mu} l_R \beta\right) Z_n
\label{nc}
\end{equation} 

where $c_W\equiv \cos\theta_W$ and $\alpha$ and $\beta$ couplings presented in the Appendix.

\section{Current Bounds}

Using the LHC7TeV data, bounds on vector doubly-charged gauge bosons of $570$~GeV has been placed on their masses \cite{Meirose:2011cs}. Moreover, analyses based on flavor changing neutral currents processes that might be applicable to this model would exclude $Z^{\prime}$ masses below 11 TeV \cite{Sanchez:2008qv}. The latter is still sensitive to the parametrization scheme used in the quark sector and also to the choose of what family of quarks transforms as an anti-4-plet. Thus we use the former as reference. 

\section{Muon Magnetic Moment}

\begin{figure}[!t]
\centering
\subfigure[\label{feyn1}]{\includegraphics[scale=0.6]{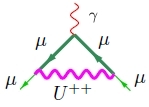}}
\subfigure[\label{feyn2}]{\includegraphics[scale=0.6]{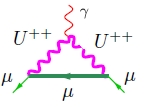}}
\subfigure[\label{feyn3}]{\includegraphics[scale=0.6]{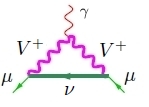}}
\subfigure[\label{feyn4}]{\includegraphics[scale=0.6]{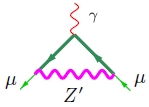}}
\caption{Diagrams that contribute to the muon magnetic moment. Two diagrams coming from doubly charged gauge bosons ({\it a} and {\it b})  , singly charged ({\it c}) and neutral gauge bosons ({\it d}).} \label{feyn}
\end{figure}

Having said that, in this work our goal is to assess whether this model accommodates the muon magnetic moment and derive $1\sigma$ limits based on current and projected sensitivities if applicable. There are new contributions to the muon magnetic moment arising in this model, namely two singly charged vector bosons $V_1$ and $V_2$ (Fig.\ref{feyn}c), a doubly charged gauge boson (Figs.\ref{feyn}a-b) and two neutral gauge bosons (Fig.\ref{feyn}d). The corrections coming from charged and neutral scalars are suppressed because their couplings with the muon are proportional to the muon mass. Hence, they are hereafter ignored. We have explicitly derived all relevant corrections to the muon magnetic moment in the Appendix. In what follows, we will show the numerical results.

\begin{figure}[!h]
\centering
\includegraphics[scale=0.8]{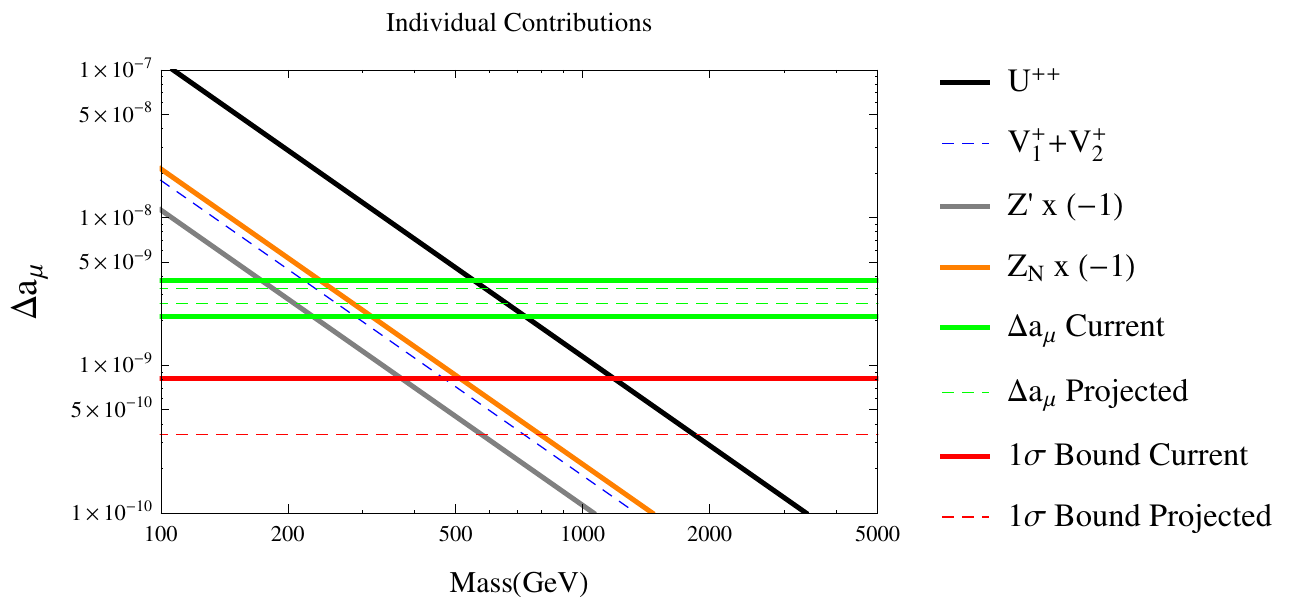}
\includegraphics[scale=0.8]{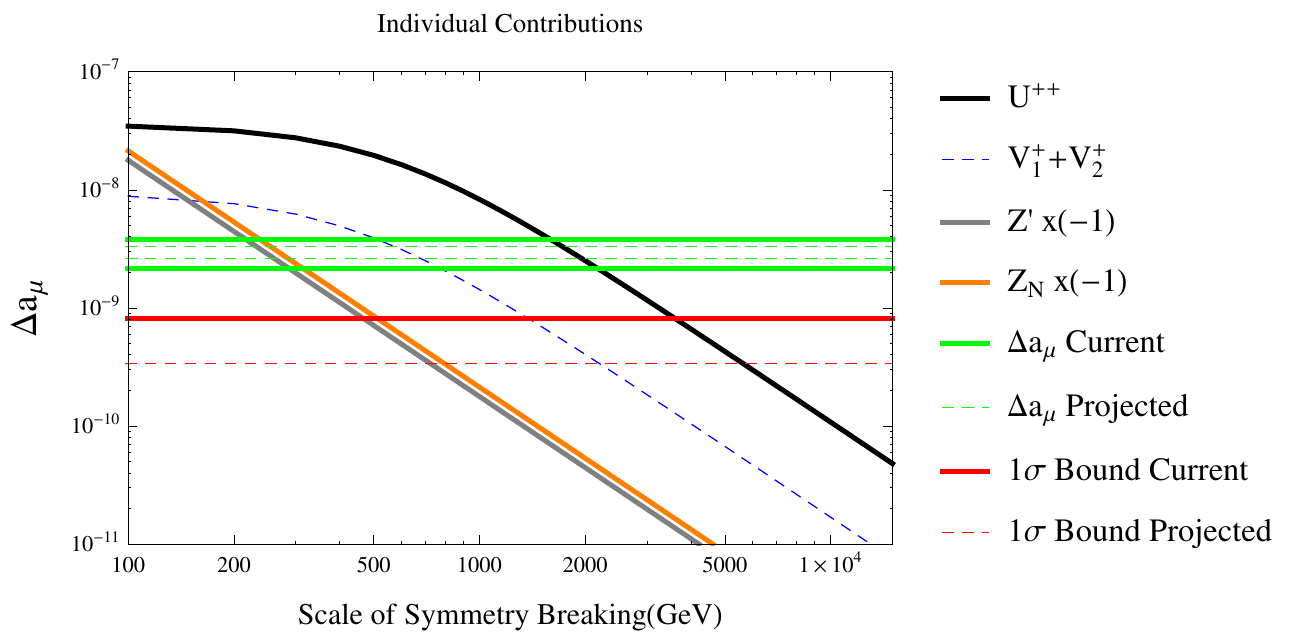}
\caption{Individual contributions to the muon magnetic moment stemming from our model as function of the particles masses (a) and the scale of symmetry breaking (b). The doubly charged boson contribution is the most relevant one by one order of magnitude, and it is positive.}
\label{fig1}
\end{figure}

In the upper panel of Fig.\ref{fig1} we exhibit the individual contributions of each one of those particles as a function of their masses. In the lower panel of Fig.\ref{fig1} we show their contributions as a function of the scale of symmetry breaking of the $SU(4)_L \otimes U(1)_N$ gauge group. We emphasize that the latter is the most reasonable way to present individual contributions, because the scale of symmetry breaking is an universal parameter in the model and it sets the masses of the particles. We conclude from Fig.\ref{fig1} that the doubly charged gauge boson ($U^{\pm \pm}$) gives rise to the most relevant contribution (see Appendix for details). The $Z^{\prime}$ and $Z_N$ correction which are proportional to $(g_{V}^{2} - 5g_{A}^2)$, due to the relative magnitude of the vector and axial couplings it turns out to be negative. The singly charged gauge bosons ($V_1^+,V_2^+$) contribution is comparable to the $Z^{\prime}$ and $Z_N$ one, but positive though. In the lower panel of Fig.\ref{fig1} the difference between their contributions is more visible, because their masses have different dependences with the scale of symmetry breaking.

\begin{figure}[!h]
\centering
\includegraphics[scale=0.8]{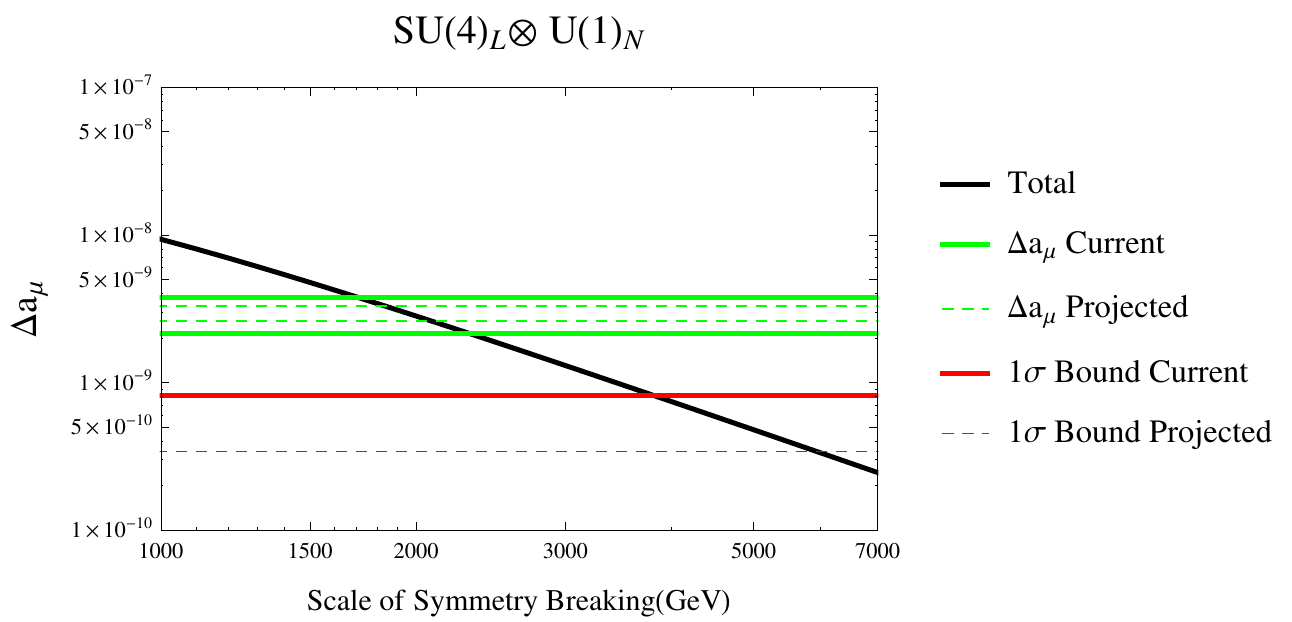}
\caption{The total contribution to the muon magnetic moment coming as function of the scale of symmetry breaking. We find it to be positive. Thus this model can accommodate the muon magnetic moment excess, and in case the anomaly is otherwise resolved with current (projected) sensitive we find that the scale of symmetry breaking must be greater than 3.5 TeV (5.5 TeV), which is the most stringent bound in the literature concerning this model.}
\label{fig2}
\end{figure}

In Fig.\ref{fig2} we have combined those individual contributions, despite the insignificance of some of them, and plotted the total contributions stemming from the model. We find that the total contribution is positive as obviously expected. Because the total contribution is positive the model can address the muon magnetic moment excess for a scale of symmetry breaking around $2$~TeV. Additionally, we may derive $1\sigma$ bounds based upon on assumption that the anomaly is otherwise resolved by any other means, using current and projected sensitivities. With current sensitivity we obtain a 3.5 TeV bound. Moreover, using the projected sensitive reported by the g-2 Fermilab experiment which is expected to take data in the near future we exclude scales of symmetry breaking smaller than 5.5 TeV. The latter is the strongest bound on the scale of symmetry breaking of this model in the literature. Further data from the LHC and g-2 experiments will shed light on the matter, but it is clear the this model is a plausible explanation to the muon magnetic moment.

\section{Conclusions}

We studied an electroweak extension of the Standard Model based on the $SU(4)_L \otimes U(1)_N$ gauge symmetry with the goal of computing the correction to the muon magnetic moment stemming from this model. We find that the doubly charged gauge boson gives by one order of magnitude the most important contribution. Interestingly, its contribution turns out to be positive. Hence we conclude that such model can address the muon magnetic moment excess with a scale of symmetry breaking of around $2$~TeV, and using current sensitivity of experiments we draw a 3.5 TeV bound on the scale of symmetry breaking of the model, in case the anomaly is otherwise resolved. Furthermore, we estimate a 5.5 TeV limit on the latter, using the project sensitive of the upcoming g-2 Fermilab experiment. We emphasize that those are the most stringent constraints in the literature found on this model.

\acknowledgements

DC is partly supported by the Brazilian National Council
for Scientific and Technological Development (CNPq) Grant
484157/2013-2. We  also would like to thank Farinaldo Queiroz and Maira Dutra for useful discussions.

\section{Appendix}

\subsection{Masses of the gauge bosons}
Coupling constant that appears in the mass of the neutral gauge bosons according to Eq.\ref{masszs}. 
\begin{mathletters}
\label{lambda}
\begin{equation}
\lambda_{n}=\frac{1}{3}\left[A+2\left(A^2+3B\right)^{\frac{1}{2}}
\cos\left(\frac{2n\pi+\Theta}{3} \right)
 \right],
\label{ln}
\end{equation}where,
\begin{equation}
A=3+4t^{2}+(7+tt^{2})a^{2}, B=-2\left[1+3t^{2}+2(4+9t^{2})a^{2}\right],
\label{ab}
\end{equation}
\begin{equation}
C=\frac{3}{32}(1+4t^2)a^2,\quad \Theta=\arccos\left[
\frac{2A^3+9AB+27C}{2(A^2+3B)^{\frac{3}{2}}}\right],
\label{c}
\end{equation}
\end{mathletters}
and we have defined $a\equiv v_1/v_2$, with $t=g^{\prime}/g$.

\subsection{Vector and axial couplings}
 
The derivation of the vector and axial couplings is a bit tedious, our results agree with Ref.\cite{Pisano:1994tf}. Defining,

\begin{equation}
Z_{n\mu}\approx x_nW^3_\mu+y_nW^8_\mu+z_nW^{15}_\mu+w_nB,
\label{autovec}
\end{equation}with
\begin{mathletters}
\label{xyzw}
\begin{equation}
x_n=-\frac{2a^2}{t}\cdot\frac{1-3t^{2}+(1-t^{2})a^{2}-(1-2t^{2})\lambda_n}{D_{n}(t,a)}\cdot w_{n},
\label{1a}
\end{equation}
\begin{eqnarray}
y_n&=&\frac{1}{\sqrt3 t}\frac{2(2+t^{2})a^{2}-10a^{4}t^{2}-\left[1+(1-4t^{2})a^{2}\right]\lambda_n}{D_{n}(t,a)}w_n,
\label{2a}
\end{eqnarray}
\begin{equation}
z_n=\frac{1}{\sqrt6t}\cdot \frac{8(2+t^{2})a^{2}+4(3+2t^{2})a^{4}-4\left[1+2(2+t^{2})a^{2}\right]\lambda_{n}+
3\lambda_{n}^{2}}{D_{n}(t,a)} \cdot w_{n},
\label{3a}
\end{equation}
\begin{equation}
w^2_n=\frac{1}{1+x_n^2/w_n^2+y_n^2/w_n^2+z_n^2/w_n^2},
\label{4a}
\end{equation}
\end{mathletters}
and finally
\begin{equation}
D_{n}(t,a)=2(7+5a^{2})-(3+13a^{2})\lambda_{n}+2\lambda_{n}^{2}.
\end{equation}

The vector and axial couplings can be derived from  the Lagrangian
 \begin{equation}
L = -\frac{g}{2C_W} \left( \bar{l_L} \gamma^{\mu} l_L \alpha  + \bar{l_R} \gamma^{\mu} l_R \beta\right) Z_n,
\end{equation}with

 \begin{equation}
\alpha=-c_W\left(-x_n+\frac{1}{\sqrt3}
y_n+\frac{1}{\sqrt6}z_n+\frac{4}{3}w_nt\right) +\frac{4}{3}c_W w_nt,\quad \beta =-\frac{3}{\sqrt6}Z_n.
\label{leptons}
\end{equation}with $Z_0=Z_N, Z_1=Z,Z_2=Z^{\prime}$. We point out that relative difference between the vector and axial couplings for $Z_N$ and $Z^{\prime}$ couplings with the muon are not sensitive to the value of a and t chosen, keeping our conclusions the same. After plugging all values we find that the vector and vector-axial couplings to be: $g_v=0.74$, $g_a=0.42$ for the $Z^{\prime}$ boson, and $g_v=1.17$, $g_a=0.63$ for the $Z_N$ gauge boson. Those values should be plugged in the Eq.(\ref{vectormuon3}) to determine the neutral gauge boson contributions to the muon magnetic moment.

\subsection{Analytical Expressions for the Muon Magnetic Moment} 

\subsubsection{Neutral Vector}
The diagram that contributes to the muon anomalous magnetic moment coming from the $Z^{\prime}$ and $Z_{N}$ particles is shown in Fig. \ref{feyn4}. The contribution is given by \cite{Queiroz:2014zfa},
\begin{eqnarray}
&&
\Delta a_{\mu} (Z^{\prime}) = \frac{m_\mu^2}{8\pi^2 M_{Z^{\prime}}^{2}}\int_0^1 dx \frac{g^2_{v9} P_{v9}(x)+ g^2_{a9} P_{a9}(x) }{(1-x)(1-\lambda^2 x) +\lambda^2 x},
\label{vectormuon1}
\end{eqnarray} where $\lambda = m_{\mu}/M_{Z^{\prime}}$ and
\begin{eqnarray}
P_{v9}(x) & = & 2 x^2 (1-x) \nonumber\\
P_{a9}(x) & = & 2 x(1-x)\cdot (x-4)- 4\lambda^2 \cdot x^3.
\label{vectormuon2}
\end{eqnarray}
These integrals simplify to give a contribution of 
\begin{equation}
\Delta a_{\mu}(Z^{\prime}) = \frac{m_{\mu}^2}{4 \pi^2 M_Z^{\prime 2}}\left(\frac{1}{3}g^2_{v9} - \frac{5}{3}g^2_{a9}\right)
\label{vectormuon3}
\end{equation}in the limit $M_{Z^{\prime}} \gg m_{\mu}$.

\subsubsection{Singly Charged Vector} 

The diagram that contributes to the muon anomalous magnetic moment coming from the
gauge bosons $V_{1,2}^{\pm}$ is shown in Fig.\ref{feyn3} which result in \cite{Queiroz:2014zfa},
\begin{eqnarray}
&&
\Delta a_{\mu} (W^{\prime}) = \frac{1}{8\pi^2}\frac{m_\mu^2}{ M_{V^+}^2 } \int_0^1 dx \frac{g_{v10}^2 \ P_{v10}(x) + g_{a10}^2 \ P_{a10} (x) }{\epsilon^2 \lambda^2 (1-x)(1-\epsilon^{-2} x) + x},
\label{vectormuon4}
\end{eqnarray}where 
\begin{eqnarray}
P_{v10}(x) & = &  2x^2(1+x-2\epsilon)+\lambda^2(1-\epsilon)^2 x(1-x)(x+\epsilon) \nonumber\\
P_{a10}(x) & = &  2x^2(1+x+2\epsilon)+\lambda^2(1+\epsilon)^2 x(1-x)(x-\epsilon),
\label{vectormuon5}
\end{eqnarray}with $\epsilon = m_{\nu}/m_{\mu}$ and $\lambda= m_{\mu}/M_{W^{\prime}}$. This simplifies to
\begin{eqnarray}
&&
\Delta a_{\mu} (W^{\prime}) = \frac{1}{4\pi^2}\frac{m_\mu^2}{ M_{W^{\prime}}^2 } \left[g_{v10}^2 \left( \frac{5}{6} - \frac{m_{\nu}}{m_{\mu}}\right)  + g_{a10}^2 \left(  \frac{5}{6} + \frac{m_{\nu}}{m_{\mu}} \right) \right]
\label{vectormuon6}
\end{eqnarray}

\subsubsection{Doubly Charged Vector} 

The diagrams that contribute to the muon anomalous magnetic moment coming from the
doubly-charged vector boson are shown in the Figs.\ref{feyn1}-\ref{feyn2} which result in \cite{Queiroz:2014zfa},

\begin{eqnarray}
&\Delta a_{\mu} (U^{\pm \pm}) &=8\times \frac{1}{8\pi^2}\left( \frac{m_\mu}{M_{U^{\pm \pm}}}\right)^2\int_0^1 dx \frac{g_{v11}^2 P_{v11}(x) + g_{a11}^2 P_{a11} (x) }{\lambda^2(1-x)^2 + x }\nonumber\\
&  & (-4)\times \frac{1}{8\pi^2}\left( \frac{m_\mu}{M_{U^{\pm \pm}}}\right)^2\int_0^1 dx \frac{g^2_{v11} P_{v11}^{\prime}(x)+ g^2_{a11} P_{a11}^{\prime} (x) }{(1-x)(1-\lambda^2 x) + \lambda^2 x},
\label{Vcontri}
\end{eqnarray}where $\lambda = m_{\mu}/M_{U^{\pm \pm}}$, and
\begin{eqnarray}
P_{v11}(x) & = & 2 x^2(x-1) \nonumber\\
P_{a11}(x) & = & 2 x^2(x+3)+4 \lambda^2 \cdot x (1-x)(x-1), \nonumber\\
P_{v11}^{\prime}(x) & = & 2 x (1-x)\cdot x \nonumber\\
P_{a11}^{\prime}(x) & = & 2 x(1-x)\cdot (x-4)- 4\lambda^2 \cdot x^3.
\end{eqnarray}
Thus the total doubly-charged vector contribution is,
\begin{eqnarray}
\Delta a_{\mu} (U^{\pm \pm})&=& \frac{ m_{\mu}^2}{\pi^2 M_{U^{\pm \pm}}^2}\left( \frac{-2}{3}g_{v11}^2 + \frac{16}{3}g_{a11}^2 \right).
\label{doublyvector}
\end{eqnarray} 

We point out that the vector current of doubly charged gauge bosons vanishes because involve identical fields in agreement with \cite{Kelso:2013zfa}. In other words, $g_{v11} \equiv 0$, and the correction from the doubly charged is thus positive.


\begin{references}
\bibitem{PDG} J. Beringer et al., Particle Data Group. Phys. Rev. D {\bf86}, 010001 (2012).

\bibitem{5loops1} T. Kinoshita and M. Nio, Phys. Rev. D {\bf73}, 013003 (2006).
\bibitem{5loops2} T. Aoyama et al., Phys. Rev. Lett. {\bf99}, 110406 (2007).
\bibitem{5loops3} T. Kinoshita and M. Nio, Phys. Rev. D {\bf70}, 113001 (2004).
\bibitem{5loops4} T. Kinoshita, Nucl. Phys. B {\bf144}, 206 (2005)(Proc. Supp). 
\bibitem{5loops5} T. Kinoshita and M. Nio, Phys. Rev. D {\bf73}, 053007 (2006).
\bibitem{5loops6} A.L. Kataev, arXiv:hep-ph/0602098 (2006).
\bibitem{5loops7} M. Passera, J. Phys. G {\bf31}, 75 (2005).

\bibitem{Hanneke:2008tm1} 
  D.~Hanneke, S.~Fogwell and G.~Gabrielse,
  Phys.\ Rev.\ Lett.\  {\bf 100}, 120801 (2008).
\bibitem{Hanneke:2008tm2} 
  G.~Gabrielse, D.~Hanneke, T.~Kinoshita, M.~Nio and B.~C.~Odom,
  Phys.\ Rev.\ Lett.\  {\bf 97}, 030802 (2006)
  [Erratum-ibid.\  {\bf 99}, 039902 (2007)].
  
\bibitem{bennet} G.W. Bennett et al., Phys. Rev. Lett. 89, 101804 (2002) (Erratum.
Phys. Rev. Lett. {\bf89}, 129903 (2002)). 
\bibitem{bennet1} G.W. Bennett et al., Phys. Rev. Lett. {\bf92}, 161802 (2004). 
\bibitem{bennet2}G.W. Bennett et al., Phys. Rev. D {\bf73}, 072003 (2006).
 
\bibitem{carey}
R. M. Carey, K. R. Lynch, J. P. Miller, B. L. Roberts, W. M.
Morse, Y. K. Semertzides, V. P. Druzhinin and B. I. Khazin
et al., FERMILAB-PROPOSAL-0989.
\bibitem{carey1} 
Andreas S. Kronfeld et al,[arXiv:1306.5009].  
  
\bibitem{Pisano:1991ee} 
  F.~Pisano and V.~Pleitez,
  Phys.\ Rev.\ D {\bf 46}, 410 (1992)
  [hep-ph/9206242].
\bibitem{Pisano:1991eee} 
  F.~Queiroz, C.~A.~de S.Pires and P.~S.~R.~da Silva,
  Phys.\ Rev.\ D {\bf 82}, 065018 (2010)
  [arXiv:1003.1270 [hep-ph]].  


\bibitem{Foot:1994ym} 
  R.~Foot, H.~N.~Long and T.~A.~Tran,
  Phys.\ Rev.\ D {\bf 50}, 34 (1994)
  [hep-ph/9402243].
\bibitem{Foot:1994ymm}  
  H.~N.~Long,
  Phys.\ Rev.\ D {\bf 53}, 437 (1996)
  [hep-ph/9504274].  

\bibitem{331Higgsportal}

D.~Cogollo, A.~X.~Gonzalez-Morales, F.~S.~Queiroz and P.~R.~Teles,
  JCAP {\bf 1411}, 002 (2014) [arXiv:1402.3271 [hep-ph]].
\bibitem{331Higgsportal1}  
P.~V.~Dong, N.~T.~K.~Ngan and D.~V.~Soa,
  arXiv:1407.3839 [hep-ph].
\bibitem{331Higgsportal2}  
  P.~V.~Dong, T.~P.~Nguyen and D.~V.~Soa,
  Phys.\ Rev.\ D {\bf 88}, no. 9, 095014 (2013)
  [arXiv:1308.4097 [hep-ph]]. 
\bibitem{331Higgsportal3}
F.~S.~Queiroz and K.~Sinha,
  Phys.\ Lett.\ B {\bf 735}, 69 (2014)
  [arXiv:1404.1400 [hep-ph]].  
\bibitem{331Higgsportal4}
 C.~A.~de S.Pires, F.~S.~Queiroz and P.~S.~Rodrigues da Silva,
  Phys.\ Rev.\ D {\bf 82}, 105014 (2010)
  [arXiv:1002.4601 [hep-ph]].
\bibitem{331Higgsportal5}
F.~S.~Queiroz, K.~Sinha and A.~Strumia,
  arXiv:1409.6301 [hep-ph].
  
\bibitem{Mizukoshi:2010ky} 
  J.~K.~Mizukoshi, C.~A.~de S.Pires, F.~S.~Queiroz and P.~S.~Rodrigues da Silva,
  Phys.\ Rev.\ D {\bf 83}, 065024 (2011)
  [arXiv:1010.4097 [hep-ph]].
\bibitem{Mizukoshi:2010ky1}
  J.~D.~Ruiz-Alvarez, C.~A.~de S.Pires, F.~S.~Queiroz, D.~Restrepo and P.~S.~Rodrigues da Silva,
  Phys.\ Rev.\ D {\bf 86}, 075011 (2012)
  [arXiv:1206.5779 [hep-ph]].
\bibitem{Mizukoshi:2010ky2}
  S.~Profumo and F.~S.~Queiroz,
  Eur.\ Phys.\ J.\ C {\bf 74}, 2960 (2014)
  [arXiv:1307.7802 [hep-ph]].
\bibitem{Mizukoshi:2010ky3}  
  A.~Alves, S.~Profumo and F.~S.~Queiroz,
  JHEP {\bf 1404}, 063 (2014)
  [arXiv:1312.5281 [hep-ph]].            
\bibitem{Mizukoshi:2010ky4}
  P.~V.~Dong, D.~T.~Huong, F.~S.~Queiroz and N.~T.~Thuy,
  arXiv:1405.2591 [hep-ph].

\bibitem{Profumo:2014mpa}
S.~Profumo and F.~S.~Queiroz,
  JCAP {\bf 1405}, 038 (2014)
  [arXiv:1401.4253 [hep-ph]].

\bibitem{Alves:2014yha} 
  A.~Alves, S.~Profumo, F.~S.~Queiroz and W.~Shepherd,
  Phys. Rev. D {\bf 90}, 115003 (2014) [arXiv:1403.5027 [hep-ph]].
\bibitem{Alves:2014yha1}
  D.~Hooper, C.~Kelso and F.~S.~Queiroz,
  Astropart.\ Phys.\  {\bf 46}, 55 (2013)
  [arXiv:1209.3015 [astro-ph.HE]].  
\bibitem{Alves:2014yha2}
  A.~X.~Gonzalez-Morales, S.~Profumo and F.~S.~Queiroz,
  arXiv:1406.2424 [astro-ph.HE].
  
\bibitem{Kelso:2013nwa} 
  C.~Kelso, C.~A.~de S. Pires, S.~Profumo, F.~S.~Queiroz and P.~S.~Rodrigues da Silva,
  Eur.\ Phys.\ J.\ C {\bf 74}, 2797 (2014)
  [arXiv:1308.6630 [hep-ph]].
\bibitem{Kelso:2013nwa1}
  D.~Hooper, F.~S.~Queiroz and N.~Y.~Gnedin,
  Phys.\ Rev.\ D {\bf 85}, 063513 (2012)
  [arXiv:1111.6599 [astro-ph.CO]].
\bibitem{Kelso:2013nwa2}
  C.~Kelso, S.~Profumo and F.~S.~Queiroz,
  Phys.\ Rev.\ D {\bf 88}, no. 2, 023511 (2013)
  [arXiv:1304.5243 [hep-ph]].
\bibitem{Kelso:2013nwa3}
  F.~S.~Queiroz,
  AIP Conf.\ Proc.\  {\bf 1604}, 83 (2014)
  [arXiv:1310.3026 [astro-ph.CO]].
\bibitem{Kelso:2013nwa4}
  F.~S.~Queiroz, K.~Sinha and W.~Wester,
  Phys. Rev. D {\bf 90}, 115009 (2014) [arXiv:1407.4110 [hep-ph].


\bibitem{Alves:2011kc} 
  A.~Alves, E.~Ramirez Barreto, A.~G.~Dias, C.~A.~de S.Pires, F.~S.~Queiroz and P.~S.~Rodrigues da Silva,
  Phys.\ Rev.\ D {\bf 84}, 115004 (2011)
  [arXiv:1109.0238 [hep-ph]].
\bibitem{Alves:2011kc1}
  A.~Alves, E.~Ramirez Barreto, A.~G.~Dias, C.~A.~de S.Pires, F.~S.~Queiroz and P.~S.~Rodrigues da Silva,
  Eur.\ Phys.\ J.\ C {\bf 73}, 2288 (2013)
  [arXiv:1207.3699 [hep-ph]].  
\bibitem{Alves:2011kc2}
  W.~Caetano, C.~A.~de S. Pires, P.~S.~Rodrigues da Silva, D.~Cogollo and F.~S.~Queiroz,
  Eur.\ Phys.\ J.\ C {\bf 73}, 2607 (2013)
  [arXiv:1305.7246 [hep-ph]].

\bibitem{othermotiv1}
D.~Cogollo, H.~Diniz, C.~A.~de S.Pires and P.~S.~Rodrigues da Silva,
  Mod.\ Phys.\ Lett.\ A {\bf 23}, 3405 (2009)
  [arXiv:0709.2913 [hep-ph]].
\bibitem{othermotiv11}
  D.~Cogollo, H.~Diniz, C.~A.~de S.Pires and P.~S.~Rodrigues da Silva,
  Eur.\ Phys.\ J.\ C {\bf 58}, 455 (2008)
  [arXiv:0806.3087 [hep-ph]].
\bibitem{othermotiv12}
  D.~Cogollo, H.~Diniz and C.~A.~de S.Pires,
  Phys.\ Lett.\ B {\bf 677}, 338 (2009) [arXiv:0903.0370 [hep-ph]].
\bibitem{othermotiv13}
  W.~Caetano, D.~Cogollo, C.~A.~de S.Pires and P.~S.~Rodrigues da Silva,
  Phys.\ Rev.\ D {\bf 86}, 055021 (2012)
  [arXiv:1206.5741 [hep-ph]].
\bibitem{othermotiv14}
  D.~Cogollo, A.~V.~de Andrade, F.~S.~Queiroz and P.~Rebello Teles,
  Eur.\ Phys.\ J.\ C {\bf 72}, 2029 (2012)
  [arXiv:1201.1268 [hep-ph]].
\bibitem{othermotiv15} 
  D.~Cogollo, F.~S.~Queiroz and P.~Vasconcelos,
  Mod.\ Phys.\ Lett.\ A {\bf 29}, 1450173 (2014) [arXiv:1312.0304 [hep-ph]].
      
\bibitem{othermotiv16}
  A.~G.~Dias and V.~Pleitez,
  Phys.\ Rev.\ D {\bf 69}, 077702 (2004)
  [hep-ph/0308037].
\bibitem{othermotiv17} 
  A.~G.~Dias, C.~A.~de S. Pires and P.~S.~R.~da Silva,
  Phys.\ Rev.\ D {\bf 68}, 115009 (2003)
  [hep-ph/0309058].
\bibitem{othermotiv18}
  A.~G.~Dias, A.~Doff, C.~A.~de S.Pires and P.~S.~Rodrigues da Silva,
  Phys.\ Rev.\ D {\bf 72}, 035006 (2005)
  [hep-ph/0503014].

\bibitem{othermotiv2}

A.~Palcu,
  arXiv:1408.6518 [hep-ph].
\bibitem{othermotiv21}
   J.~E.~C.~Montalvo, C.~A.~M.~Cruz, R.~J.~G.~Ramirez, G.~H.~R.~Ulloa, A.~I.~R.~Mendoza and M.~D.~Tonasse,
  arXiv:1408.5944 [hep-ph].
\bibitem{othermotiv22} 
  J.~E.~C.~Montalvo, R.~J.~G.~Ramírez, G.~H.~R.~Ulloa, A.~I.~R.~Mendoza and Tonasse. M.D.,
  arXiv:1311.0845 [hep-ph].
\bibitem{othermotiv23}
   J.~E.~Cieza Montalvo, R.~J.~Gil Ramírez, G.~H.~Ramírez Ulloa, A.~I.~Rivasplata Mendoza and M.~D.~Tonasse,
  Phys.\ Rev.\ D {\bf 88}, no. 9, 095020 (2013)
  [arXiv:1205.4042 [hep-ph]].
\bibitem{othermotiv24}  
  J.~E.~C.~Montalvo, G.~H.~R.~Ulloa and M.~D.~Tonasse,
  Eur.\ Phys.\ J.\ C {\bf 72}, 2210 (2012)
  [arXiv:1205.3822 [hep-ph]].     
\bibitem{othermotiv25}
  A.~Palcu,
  arXiv:1102.1180 [hep-ph].
\bibitem{othermotiv26}
  A.~Palcu,
  arXiv:0802.3588 [hep-ph].
\bibitem{othermotiv27}
  A.~Palcu,
  Mod.\ Phys.\ Lett.\ A {\bf 23}, 387 (2008)
  [arXiv:0801.0036 [hep-ph]].
\bibitem{othermotiv28}
  C.~Alvarado, R.~Martinez and F.~Ochoa,
  Phys.\ Rev.\ D {\bf 86}, 025027 (2012)
  [arXiv:1207.0014 [hep-ph]].
\bibitem{othermotiv29}
  A.~E.~Carcamo Hernandez, R.~Martinez and F.~Ochoa,
  Phys.\ Rev.\ D {\bf 87}, no. 7, 075009 (2013)
  [arXiv:1302.1757 [hep-ph]].
\bibitem{othermotiv210}
  R.~Martinez, F.~Ochoa and J.~P.~Rubio,
  Phys.\ Rev.\ D {\bf 89}, 056008 (2014)
  [arXiv:1303.2734 [hep-ph]].
\bibitem{othermotiv211}
  R.~Martinez and F.~Ochoa,
  Phys.\ Rev.\ D {\bf 90}, 015028 (2014)
  [arXiv:1405.4566 [hep-ph]].
\bibitem{othermotiv212}
  J.~G.~Duenas, N.~Gutierrez, R.~Martinez and F.~Ochoa,
  Eur.\ Phys.\ J.\ C {\bf 60}, 653 (2009).

\bibitem{Pisano:1994tf} 
  F.~Pisano and V.~Pleitez,
  Phys.\ Rev.\ D {\bf 51}, 3865 (1995)
  [hep-ph/9401272].
   
\bibitem{Kelso:2013zfa} 
  C.~Kelso, P.~R.~D.~Pinheiro, F.~S.~Queiroz and W.~Shepherd,
  Eur.\ Phys.\ J.\ C {\bf 74}, 2808 (2014)
  [arXiv:1312.0051 [hep-ph]].
\bibitem{Kelso:2013zfa1}
  N.~A.~Ky, H.~N.~Long and D.~V.~Soa,
  Phys.\ Lett.\ B {\bf 486}, 140 (2000)
  [hep-ph/0007010].
\bibitem{Kelso:2013zfa2}
   C.~Kelso, H.~N.~Long, R.~Martinez and F.~S.~Queiroz,
  arXiv:1408.6203 [hep-ph].        
  
\bibitem{Palcu:2009ks} 
  A.~Palcu,
  Mod.\ Phys.\ Lett.\ A {\bf 24}, 1247 (2009)
  [arXiv:0902.1301 [hep-ph]].
\bibitem{Palcu:2009ks1}
  A.~Palcu,
  Mod.\ Phys.\ Lett.\ A {\bf 24}, 2589 (2009)
  [arXiv:0908.1636 [hep-ph]].
\bibitem{Palcu:2009ks2}
   A.~G.~Dias, P.~R.~D.~Pinheiro, C.~A.~de S.Pires and P.~S.~Rodrigues da Silva,
  Annals Phys.\  {\bf 349}, 232 (2014)
  [arXiv:1309.6644 [hep-ph]].

\bibitem{Liu:1994rx} 
  J.~T.~Liu,
  Phys.\ Rev.\ D {\bf 50}, 542 (1994)
  [hep-ph/9312312].
    
  
\bibitem{Meirose:2011cs} 
  B.~Meirose and A.~A.~Nepomuceno,
  Phys.\ Rev.\ D {\bf 84}, 055002 (2011)
  [arXiv:1105.6299 [hep-ph]].
  
\bibitem{Sanchez:2008qv} 
  L.~A.~Sanchez, L.~A.~Wills-Toro and J.~I.~Zuluaga,
  Phys.\ Rev.\ D {\bf 77}, 035008 (2008)
  [arXiv:0801.4044 [hep-ph]].

\bibitem{Queiroz:2014zfa} 
  F.~S.~Queiroz and W.~Shepherd,
  Phys.\ Rev.\ D {\bf 89}, 095024 (2014)
  [arXiv:1403.2309 [hep-ph]].      
\end{references}
\end{document}